\documentclass[%
 reprint,prb,
superscriptaddress
]{revtex4-1}

\usepackage{bm}%
\usepackage[colorlinks=true,linkcolor=blue]{hyperref}%
\expandafter\ifx\csname package@font\endcsname\relax\else
 \expandafter\expandafter
 \expandafter\usepackage
 \expandafter\expandafter
 \expandafter{\csname package@font\endcsname}%
\fi
\usepackage{graphicx}
\usepackage{dcolumn}

\begin{document}

\preprint{APS/123-QED}

\title{Dirac wave transmission in L\'evy disordered systems}

\author{Jonas R. F. Lima}
\email{jonas.lima@ufrpe.br}
\affiliation{Departamento de F\'{\i}sica, Universidade Federal Rural de Pernambuco, 52171-900, Recife, PE, Brazil }

\author{Luiz Felipe C. Pereira}
\email{pereira@fisica.ufrn.br}
\affiliation{Departamento de F\'{\i}sica, Universidade Federal do Rio Grande do Norte, 59078-970, Natal, RN, Brazil}

\author{Anderson L. R. Barbosa}
\email{anderson.barbosa@ufrpe.br}
\affiliation{Departamento de F\'{\i}sica, Universidade Federal Rural de Pernambuco, 52171-900, Recife, PE, Brazil }

\date{\today}

\begin{abstract}
We investigate the propagation of electronic waves described by the Dirac equation subject to a L\'evy-type disorder distribution. Our numerical calculations, based on the transfer matrix method, in a system with a distribution of potential barriers show that it presents a phase transition from anomalous to standard to anomalous localization as the incidence energy increases. In contrast, electronic waves described by the Schr\"odinger equation do not present such transitions. Moreover, we obtain the phase diagram delimiting anomalous and standard localization regimes, in the form of an incidence angle versus incidence energy diagram, and argue that transitions can also be characterized by the behavior of the dispersion of the transmission. We attribute this transition to an abrupt reduction in the transmittance of the system when the incidence angle is higher than a critical value which induces a decrease in the transmission fluctuations.
\end{abstract}

\pacs{Valid PACS appear here}
\maketitle


\section{Introduction}

Anderson localization (also known as strong localization) is a remarkable physical phenomenon characterized by a complete suppression of wave diffusion in a disordered media due to destructive quantum interference [\onlinecite{PhysRev.109.1492}].
At first, the phenomenon was successfully proposed  for electronic transport problems, introducing the quantum mechanical interpretation of electronic motion in a disordered device [\onlinecite{PhysicsToday.62}].
A similar phenomenon was later observed for light waves, which became known as the Anderson localization of light [\onlinecite{JOUR}]. 
An anomalous localization behavior, different from the standard Anderson localization, has been obtained when the probability density of the disorder distribution has a long tail, as in the case of L\'evy distributions [\onlinecite{0295-5075-92-5-57014,PhysRevB.24.5698,0305-4470-36-12-322,PhysRevB.85.235450,PhysRevB.88.205414,PhysRevLett.117.046603}]. 
In the last three decades several stochastic phenomena have been described by the statistics of L\'evy distributions, such as human mobility [\onlinecite{Song1018}], fluid dynamics [\onlinecite{PhysRevLett.78.3864,PhysRevLett.71.3975,PhysRevE.95.032315}], photons [\onlinecite{Nature453,PhysRevE.91.032112,PhysRevB.98.235144,PhysRevA.85.035803,vanLoevezijn:96,2040-8986-17-10-105601,doi:10.1080/09500340.2017.1310317}], random lasers [\onlinecite{PhysRevA.91.043827,NatureCommunications8}], free-standing graphene membranes [\onlinecite{PhysRevLett.117.126801}], and more recently, electronic transport [\onlinecite{0295-5075-92-5-57014,PhysRevB.24.5698,0305-4470-36-12-322,PhysRevB.85.235450,PhysRevB.88.205414,PhysRevLett.117.046603,0957-4484-28-13-134001,PhysRevB.97.075417,PhysRevE.96.062141,PhysRevE.93.012135}]. 
These phenomena provide a venue to a deeper understanding of electronic localization.

{A L\'evy distribution is characterized by a probability density $\rho(w)$ of a random variable $w$, which has a power-law tail [\onlinecite{Nature453,PhysRevE.96.062141,PhysRevB.98.235144}]. The probability density is given by $\rho(w) \propto 1/w^{1+\alpha}$, where $0 < \alpha < 2$. If $0 < \alpha < 1$, the first and second moments of $\rho(w)$ diverge because of heavy-tails, while for $1 \le \alpha < 2$ only the second moment diverges. In particular, for $\alpha = 0.5$, the so-called L\'evy distribution is analytical and given by $\rho(w) = \left(2\pi\right)^{-1/2} w^{-3/2}\exp{\left(-1/2w\right)}$, while for other values of $\alpha$ the distribution can be obtained numerically [\onlinecite{LIANG2013242}].}

{The localization of classical waves in weakly scattering one-dimensional L\'evy lattices was studied recently [\onlinecite{PhysRevE.91.032112,PhysRevB.98.235144}]. It was proven that the localization length ($\xi$) of long wavelengths is proportional to the power of the wavelength, $\xi\propto \lambda ^\alpha$, for $1 < \alpha < 2$ and that it has a transcendental behavior, $\xi\propto \lambda^2/\ln\left(\lambda\right)$, for $\alpha = 2$. However, for $\alpha > 2$, the localization length goes to typical Anderson localization which is given by $\xi\propto \lambda^2$.}

From the electronic transport point of view, a Schr\"odinger electronic wave submitted to typical one-dimensional disorder (Anderson model) shows standard localization, which means that the average transmission decays exponentially with the system length $L$ [\onlinecite{Mello,RevModPhys.69.731,PhysRevB.22.3519}],
\begin{equation}
\left\langle T \right\rangle \propto \exp{\left(-\frac{L}{2 \ell}\right)},
\label{Texp}
\end{equation}
where $\ell$ is the mean free path. 
Meanwhile the average of minus the logarithm of the transmission increases linearly with $L$,
\begin{equation}
\left\langle - \ln T \right\rangle = \frac{L}{\ell}.
\label{lnL}
\end{equation}
Eq. (\ref{lnL}) is used to obtain the mean free path from experiments in disordered one-dimensional devices [\onlinecite{RevModPhys.69.731,PhysRevB.98.155407}].
However, if the Schr\"odinger electronic wave is submitted to an one-dimensional L\'evy-type disorder, standard localization is no longer observed. 
In this case, Eqs. (\ref{Texp}) and (\ref{lnL}) become, respectively [\onlinecite{0295-5075-92-5-57014,PhysRevA.85.035803,PhysRevLett.113.233901}]
\begin{equation}
\left\langle T \right\rangle \propto L^{-\alpha},
\label{TL}
\end{equation}
\begin{equation}
\left\langle - \ln T \right\rangle \propto L^{\alpha},
\label{lnLa}
\end{equation}
where $\alpha$ is the exponent of the power-law tail in the  L\'evy distribution. 
The effects of the L\'evy-type disorder are stronger in the range $0<\alpha<1$.
The behavior described by Eqs. (\ref{TL}) and (\ref{lnLa}) is known as anomalous localization.

\begin{figure}
\centering
\includegraphics[width=0.9\linewidth]{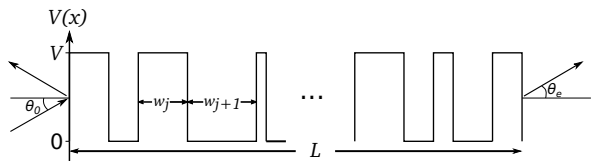}
\caption{The propagation of Dirac electronic wave along the sequence of barriers and wells with thicknesses $w_j$ and $w_{j+1}$ following a L\'evy-type distribution.}
\label{potential}
\end{figure}

Recent investigations of electronic transport in Dirac systems identified characteristics that could give rise to anomalous localization. 
For instance, in Ref. [\onlinecite{PhysRevB.88.205414}] the authors show  that typical disorder in graphene nanoribbons could originate either standard or anomalous localization depending on the type of ribbon edge.
Furthermore, in Ref. [\onlinecite{PhysRevLett.117.046603}] the transport properties of graphene with anisotropically distributed on-site impurities were investigated, showing that the system reveals L\'evy-flight transport in the stripe direction, which leads to a conductivity that increases with the square root of the system length.  
However, as far as we know [\onlinecite{1,2,3,4,5,6,7,8,LIMA2018105,0953-8984-29-15-155303,9,10,11,12,13}], there are no studies about the effects of a specific L\'evy-type disorder over a Dirac electronic wave. 
Hence, an issue remains open: what type of localization will manifest in the case of Dirac electronic waves subject to a L\'evy-type disorder?

Here we address this question by investigating the transmission of Dirac electronic waves subject to a L\'evy-type disorder, as schematically shown in Fig. \ref{potential}. 
We show that the system presents a phase transition from anomalous to standard to anomalous localization as incidence energy increases, in contrast to Schr\"odinger electronic waves which do not present  such transitions [\onlinecite{0295-5075-92-5-57014,PhysRevA.85.035803,PhysRevLett.113.233901}]. 
Moreover, we obtain the phase diagram delimiting anomalous and standard localization regimes, in the form of an incidence angle versus incidence energy diagram. 
Our numerical results have been obtained by the transfer matrix method and compared with Eqs. (\ref{Texp})-(\ref{lnLa}). 
{We believe that our results could in principle be achieved following recent experiments in graphene devices and superlattices [\onlinecite{PhysRevB.89.115421,PhysRevB.97.195410,doi:10.1063/1.4807888,10.1038/ncomms3342,KrishnaKumar181,PhysRevLett.121.036802, PhysRevX.6.041020,PhysRevB.97.075434}].}

\section{Model and Methodology}

With the aim of connecting our results with experiments, we analyze the electronic transport properties of graphene in the presence of a L\'evy-type disorder. The effective Dirac equation for fermions moving in the presence of one-dimensional potential barriers in graphene is given by
\begin{eqnarray}
-i\hbar v_F\left(\sigma_x \partial_x +\sigma_y \partial_y\right)\psi  = [E-V(x)] \psi,
\label{diraceq}
\end{eqnarray}
where $\sigma_i$ are the Pauli matrices, $v_F$ is the Fermi velocity 
{and the spinor $\psi=(\psi_A, \psi_B)^T$, with $A$ and $B$ representing the two polarizations of the pseudospin which correspond to two graphene sublattices.} 
The electrostatic potential $V(x)$ is piecewise constant, which alternates between two values, $V(x)=V$ and $0$, as depicted in Fig. \ref{potential}. 
The width of the regions $w_j$, with and without potential barriers, follows a L\'evy-type distribution. 
A similar experimental setup can, at least in principle, be achieved following recent experiments in graphene superlattices [\onlinecite{PhysRevB.89.115421,PhysRevB.97.195410,doi:10.1063/1.4807888,10.1038/ncomms3342,KrishnaKumar181,PhysRevLett.121.036802}].


We employ the transfer matrix method to calculate the transmittance directly. Since $V(x)$ is constant inside the $j$th region, we can write $\psi(x,y)=e^{-ik_yy}\psi(x)$, where $\psi(x)=(\psi_A(x), \psi_B(x))^T$, and obtain 
\begin{equation}
\frac{d^2\psi_{A,B}}{dx^2} + (k^2_j - k^2_y) \psi_{A,B} = 0  ,
\end{equation}
where $k_j = (E-V_j)/(\hbar v_F)$ is the wave vector inside of the $V_j$ barrier. The subscript $j$ denotes the regions of the system, $j = 0,1,2,...,N,e$, where $j=0$ is the incident region while $j=e$ the exit region. Note that $N$ is not a fixed number. It depends on the L\'evy distribution.

Following Ref. [\onlinecite{Yao10}], we obtain the transfer matrix connecting the wave function $\psi(x)$ at $x$ and $x+\Delta x$ in the $j$th barrier, which is given by
\begin{equation}
M_j(\Delta x, E, k_y)=\left(
\begin{array}{cc}
\frac{\cos(q_j\Delta x - \theta_j)}{\cos\theta_j} & i\frac{\sin(q_j\Delta x)}{\cos\theta_j} \\
i\frac{\sin(q_j\Delta x)}{\cos\theta_j} & \frac{\cos(q_j\Delta x + \theta_j)}{\cos\theta_j}
\end{array} \right) \; ,
\end{equation}
where $q_j$ is the $x$ component of the wave vector  given by $q_j = \sqrt{k_j^2-k_y^2}$ for $k_j^2>k_y^2$, otherwise $q_j = i\sqrt{k_y^2-k_j^2}$. The term $\theta_j$ is the angle between the $x$ component of the wave vector, $q_j$, and the wave vector, $k_j$, $\theta_j = \arcsin (k_y/k_j)$. Hence, the transfer matrix connecting incident and exit wave functions is given by $X = \prod_{j=1}^{N} M_j(w_j, E, k_y)$, while the transmission coefficient is given by
\begin{equation}
t(E, k_y) = \frac{2\cos \theta_0}{(x_{22}e^{-i\theta_0}+x_{11}e^{-i\theta_e})-x_{12}e^{i(\theta_e-\theta_0)}-x_{21}} ,
\label{tc}
\end{equation}
where $x_{mn}$ are the matrix elements of $X$ and $\theta_0 (\theta_e)$ is the incidence (exiting) angle.

\section{Results and Discussion}

\begin{figure}
\centering
\includegraphics[width=0.9\linewidth]{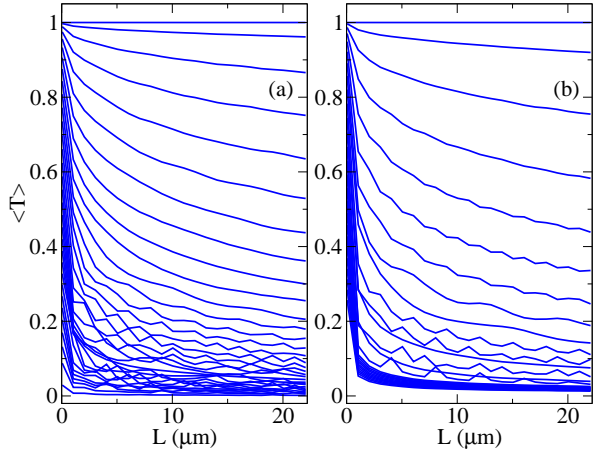}
\caption{Average transmission probability $\left\langle T \right\rangle$ as a function of system length $L$ for Dirac heterostructures with L\'evy-type distribution of potential barriers with $V = 50$ meV. Incidence angle increases from top $\theta = 0$ to bottom $\pi/2$, and incidence energies are (a) $E = 20$ meV and (b) $E = 30$ meV.}
\label{Imagem2}
\end{figure}

In Fig. \ref{Imagem2}, we present the average transmission probability $\left\langle T \right\rangle=\left\langle tt^\dagger \right\rangle$ as a function of system length $L$ for a disordered graphene superlattice with L\'evy-type potential barrier distribution characterized by $\alpha = 0.5$. The energy of the barriers is $V=50$ meV. 
The average was calculated from $10^4$ realizations for two different values of incidence energy, $E=20$ and $30$ meV, and varying the incidence angle $\theta$ from zero to $\pi/2$ by $0.05$ radians at each step. 
In the case of normal incidence the charge carriers do not feel the electrostatic potential barriers due to Klein tunneling [\onlinecite{Katsnelson}], and we obtain $\left\langle T \right\rangle=1$ for $\theta = 0$ in both cases, as expected. 
However, when $\theta$  increases the potential barrier distribution becomes relevant to the transport properties, and $\left\langle T \right\rangle$ tends to decrease as a power law, Eq. (\ref{TL}), instead of the exponential decay in Eq. (\ref{Texp}). 
This behavior indicates the onset of anomalous localization in Dirac materials introduced by a L\'evy-type disorder distribution, represented here by the distribution of potential barriers.

\begin{figure}
\centering
\includegraphics[width=0.9\linewidth]{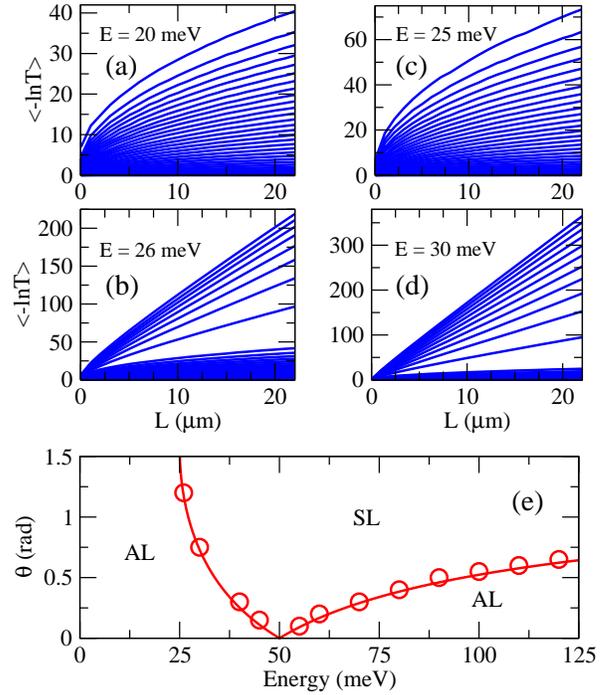}
\caption{Top panels, (a) to (d), show $\left\langle -\ln T \right\rangle$ as a function of length for Dirac heterostructures with L\'evy-type distribution of potential barriers with $V = 50$ meV. Incidence angle increases from bottom $\theta = 0$ to top $\pi/2$, and incidence energies are $E=20, 25, 26$ and $30$ meV. 
Bottom panel (e) shows the phase diagram in terms of incidence angle $\theta$ and incidence energy $E$, dividing anomalous (AL) and ``standard'' localization (SL) regimes.}
\label{Imagem3}
\end{figure}

The top panels of Fig. \ref{Imagem3} show $\left\langle -\ln T \right\rangle$ as a function of $L$ for various incidence energies, $E=20, 25, 26$ and $30$ meV, and varying the incidence angle $\theta$ from zero to $\pi/2$ in increments of $0.05$ radians.
When $E$ is equal to or smaller than half of the potential barrier energy ($E \le V/2 = 25$ meV), $\left\langle -\ln T \right\rangle$ increases as a power law in accordance with Eq. (\ref{lnLa}), which characterizes an anomalous localization behavior. 
However, when $E$ is greater than half of the potential barrier energy ($E > V/2 = 25$ meV), $\left\langle -\ln T \right\rangle$ shows a transition in localization behavior as the incidence angle increases. 

The bottom panel of Fig. \ref{Imagem3} presents the phase diagram in terms of incidence angle $\theta$ and incidence energy $E$. 
The continuous line is the critical incidence angle given by $\theta_c=\arcsin|1-V/E|$ \onlinecite{}, while the data points were obtained from our numerical calculations.
The critical incidence angle $\theta_c$ defines a transition in localization behavior, and shows a minimum value when the incidence energy equals the energy of the potential barriers $E=V=50$ meV. 
The phase diagram shows two different localization regimes: anomalous localization (AL) and ``standard'' localization (SL).
In the AL regime, $\left\langle -\ln T \right\rangle$ increases as a power law as described by Eq. (\ref{lnLa}), while in the SL regime it increases linearly with $L$, as described by Eq. (\ref{lnL}).
Notice that for $E \le V/2 = 25$ meV, the system is always in the AL regime for any incidence angle in the interval $0 \le \theta \le \pi/2$.

\begin{figure}
\centering
\includegraphics[width=0.9\linewidth]{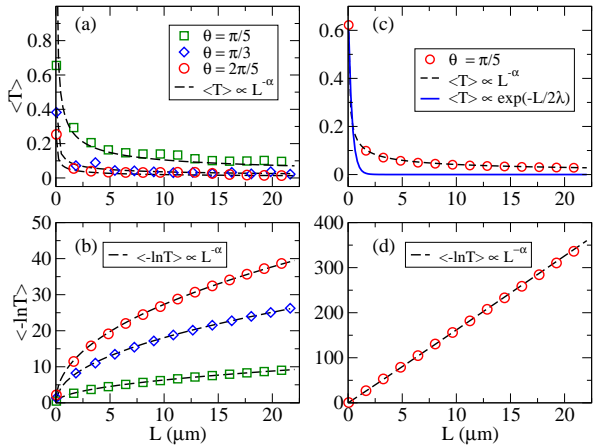}
\caption{Average transmission, (a) and (c), and $\left\langle -\ln T \right\rangle$, (b) and (d), as a function of system length for Dirac heterostructures with L\'evy-type barrier distribution for $\alpha = 0.5$. In (a) and (b) we have $E=25$ meV ($E<V$). The dashed lines are fitted using Eqs. (\ref{TL}) and (\ref{lnLa}). In (c) and (d) we have $E=60$ meV ($E>V$) and $\theta=\pi/5$. The dashed lines are fitted using Eqs. (\ref{TL}) and (\ref{lnL}), whereas the straight line is fitted with Eq. (\ref{Texp}).}
\label{Imagem4}
\end{figure}

To better understand the transition in localization behavior, we  analyze the AL and SL regions in detail. 
The former is shown on the left-hand side of Fig. \ref{Imagem4} while the latter is shown on its right-hand side. 
Fig. \ref{Imagem4} (a) shows $\left\langle T \right\rangle$ as a function of $L$ for three incidence angles $\theta=\pi/5,\pi/3$ and $2\pi/5$ and incidence energy fixed at $E = 25$ meV. 
All curves can be fitted by a power law decay, Eq. (\ref{TL}), with $\alpha=0.5$ (dashed lines). 
In addition, Fig. \ref{Imagem4} (b) shows $\left\langle -\ln T \right\rangle$ as a function of length $L$ with the same parameters. Again, the numerical data is better fitted by a power law as in Eq. (\ref{lnLa}) with $\alpha=0.5$ (dashed lines). 
This behavior indicates that Dirac carriers in the AL region feel the tunneling potential barriers with L\'evy-type distribution the same way that Schr\"odinger particles do. 
Furthermore, this behavior is valid for all values in the range $0 < \alpha <1$.

In Fig. \ref{Imagem4} (c) we present $\left\langle T \right\rangle$ as a function of length $L$ for $\theta=\pi/5$ and $E = 60$ meV.
In this case, we attempted to fit the numerical data with an exponentially decaying function as in Eq.(\ref{Texp}), and a power law decay as in Eq. (\ref{TL}). 
According to Fig. \ref{Imagem4} (c), only the power law decay is capable of describing the dependence of $\left\langle T \right\rangle$ with $L$. 
Therefore, the power law decay of $\left\langle T \right\rangle$ for $E>V$ agrees with AL behavior observed for $E<V$.
Nonetheless, Fig. \ref{Imagem4} (d) shows the dependence of $\left\langle -\ln T \right\rangle$ on $L$ with the same parameters, where we observe a linear dependence as predicted by Eq. (\ref{lnL}) (dashed lines). 
Therefore, the linear increase of $\left\langle -\ln T \right\rangle$ indicates a SL behavior instead of AL.

\begin{figure}
\centering
\includegraphics[width=0.9\linewidth]{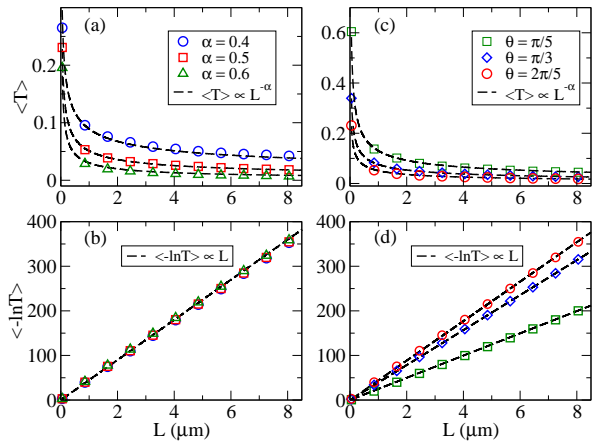}
\caption{Average transmission, (a) and (c), and $\left\langle -\ln T \right\rangle$, (b) and (d), as a function of $L$ for Dirac heterostructures with L\'evy-type barrier distribution. In (a) and (b) $\theta=2\pi/5$, while in (c) and (d) $\alpha = 0.5$.
The dashed lines are fitted using Eqs. (\ref{TL}) and (\ref{lnL}) for top and bottom panels respectively. Incidence energy is fixed $E=60$ meV.}
\label{Imagem5}
\end{figure}

In order to verify if this localization behavior is a general result for the SL region, we now analyze the data in Fig. \ref{Imagem5}. 
In Fig. \ref{Imagem5} (a) and (b), we fix the incidence angle $\theta=2\pi/5$ and consider $\alpha=0.4,0.5$ and $0.6$.  Meanwhile in Fig. \ref{Imagem5} (c) and (d), we fix $\alpha=0.5$ and change the incidence angle $\theta=\pi/5,\pi/3$ and $2\pi/5$. In all cases, the incidence energy is fixed at $E = 60$ meV. 
Fig. \ref{Imagem5}, shows that $\left\langle T \right\rangle$ presents a power law decay, while $\left\langle -\ln T \right\rangle$ increases linearly with $L$. 
This means that the SL region of the phase diagram is not a proper standard localization, in the sense of Anderson localization, because $\left\langle T \right\rangle$ is not  described by an exponential decay.
However, $\left\langle -\ln T \right\rangle$ is indeed described by a linear increase as it would be in the case of Anderson localization.

We remark that a system with carriers described by the Schr\"odinger equation also presents an anomalous to standard localization transition as a function of $\alpha$, when $\alpha \ge 1$ [\onlinecite{PhysRevA.85.035803}]. 
This is reasonable since increasing $\alpha$ is equivalent of increasing the density of disorder in the system. 
When $\alpha \ge 2$ the Schr\"odinger system presents typical Anderson localization. 
{A similar behavior occurs in classical waves: for $1 < \alpha \le 2$ the localization is anomalous while for $\alpha \ge 2$ it is standard  [\onlinecite{PhysRevB.98.235144}]}.
In contrast, our results show that a system with Dirac carriers presents a transition in localization behavior as a function of energy, with no change for $0 < \alpha < 1$, in other words, without change in the disorder density.

\begin{figure}
\centering
\includegraphics[width=0.9\linewidth]{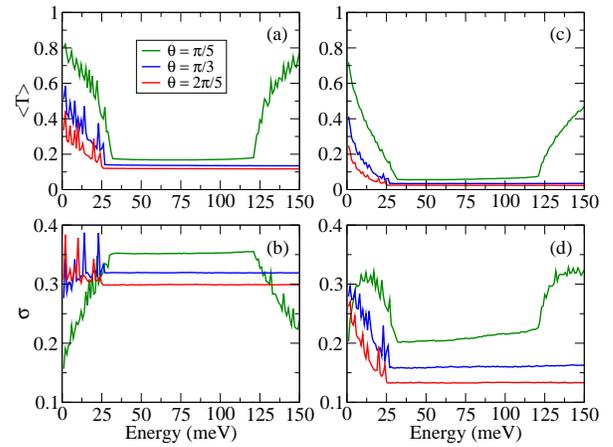}
\caption{Average transmission and its standard deviation as a function of incidence energy for a system length $L=5$ $\mu$m. In (a) and (b) we have  $\alpha=0.25$, while in (c) and (d)  $\alpha=0.50$. The incidence angle takes values $\theta=\pi/5,\pi/3$ and $2\pi/5$.}
\label{Imagem6}
\end{figure}

Finally, in Fig. \ref{Imagem6} we analyze $\left\langle T \right\rangle$, (a) and (c), and its standard deviation $\sigma = \sqrt{ \langle T^2 \rangle - \langle T \rangle^2}$, (b) and (d),  as a function of incidence energy. 
We take $L=5$ $\mu$m, and $\alpha=0.25$ in (a) and (b), while $\alpha=0.5$ in (c) and (d).
Incidence angles are $\theta=\pi/5,\pi/3$ and $2\pi/5$. 
In accordance with the phase diagram of Fig. \ref{Imagem3}, when $\theta=\pi/5$, $\left\langle T \right\rangle$ shows two transitions in localization behavior. 
First, from AL to SL at $E\approx 32$ meV, and a second one  from SL to AL at $E\approx 122$ meV. 
In the case of $\theta=\pi/3$ and $2\pi/5$, we only observe one transition from AL to SL at $E\approx 25$ meV. 
In fact, the transition from SL to AL will also occur in these cases, but at very high incidence energies.

In all cases we notice that $\left\langle T \right\rangle$ is reduced to a  small constant value when the system is in the SL regime. 
This is due to quasiparticles with incidence angle equal to $\theta_c$ being transmitted at $\pi/2$ in the potential barrier region [\onlinecite{theta}]. 
For incidence angles higher then $\theta_c$, the incident plane wave becomes an evanescent one, reducing abruptly the transmittance. 
In our system, this reduction in $\left\langle T \right\rangle$ also causes a reduction in its fluctuations, as shown in the bottom panels of Fig. \ref{Imagem6}. 
The transition in localization behavior then becomes clearer, since the AL is a consequence of the large transmission fluctuations which appear in systems with a L\'evy-type disorder. 
When these fluctuations are suppressed, which in our case happens for incidence angles higher than $\theta_c$, the system stays in the SL regime.

\section{Conclusion}

In summary, we investigated the propagation of electronic waves described by the Dirac equation in a quasi-one-dimensional system subject to a disorder distribution.
We performed numerical calculations based on the transfer matrix method in a system with a L\'evy-type distribution of potential barriers.
We have shown that the system presents a transition from anomalous to standard to anomalous localization as the incidence energy increases.
In  contrast to electronic waves described by the  Schr\"odinger  equation, which do not present such transitions.  
The phase diagram delimiting the anomalous and standard localization regimes, as a function of incidence angle and incidence energy, was obtained.
We have shown that the transitions can also be characterized by the behavior of the dispersion of the transmission.
Finally, we attribute the transitions in localization regime to an abrupt decay in the transmittance of the system which occurs when the incidence angle is higher than $\theta_c$, and induces a decrease in the transmission fluctuations.
We believe that our results could, at least in principle, be achieved following recent experiments in graphene superlattices [\onlinecite{PhysRevB.89.115421,PhysRevB.97.195410,doi:10.1063/1.4807888,10.1038/ncomms3342,KrishnaKumar181,PhysRevLett.121.036802}]. 
{For instance, recent experiments produced and characterized graphene devices with length in the  3 -- 5 $\mu$m range  [\onlinecite{PhysRevX.6.041020,PhysRevB.97.075434}], which could enable a possible experimental verification of our results.} 

\acknowledgements
This work was partially supported by Brazillian angencies Conselho Nacional de Desenvolvimento Cient\'ifico e Tecnológico (CNPq), Coordena\c{c}\~ao de Aperfei\c{c}oamento de Pessoal de N\'ivel Superior (CAPES), and Funda\c{c}\~ao de Amparo \`a Ci\^encia e Tecnologia de Pernambuco (FACEPE).

\bibliography{ref}

\end{document}